\documentclass[aps,superscriptaddress,preprintnumbers,twocolumn,floatfix,nofootinbib]{revtex4-2} 
\pdfoutput=1

\usepackage{amsfonts}
\usepackage{amsmath}
\usepackage{amssymb}
\usepackage{bm}
\usepackage{dcolumn}
\usepackage{graphicx}
\usepackage[latin1]{inputenc}
\usepackage{latexsym}
\usepackage{rotating}
\usepackage{hyperref}
\usepackage{color}
\usepackage{tensor}

\newcommand{\del}{\partial}

\def\nn{\nonumber} 

\def\a{\alpha} 
\def\b{\beta}\def\g{\gamma} 
 \def\d{\delta} 
    \def\k{\kappa}
\def\l{\lambda} \def\L{\Lambda}  \def\m{\mu}
    \def\r{\rho}
\def\s{\sigma}

\def\cA{{\cal A}}  \def\cC{{\cal C}} 
   
 \def\cH{{\cal H}}  
 \def\cK{{\cal K}}  
\def\cM{{\cal M}}  \def\cO{{\cal O}} 
  \def\cR{{\cal R}}

 \def\cZ{{\cal Z}}

\def\R{{\mathbb R}} \def\C{{\mathbb C}} 

 \def\one{\mbox{1 \kern-.59em {\rm l}}}

\def\mso{\mathfrak{so}}

\newcommand{\Tr}{\mathrm{Tr}}

\newcommand{\End}{\mathrm{Mat}}

\def\Tr{\mbox{Tr}}

\newcommand{\eq}[1]{(\ref{#1})}

\begin{document}

\preprint{UWTHPh-2021-17}

\title{Gravity as a Quantum Effect on Quantum Space-Time}

\author{Harold C. Steinacker}
\email{harold.steinacker@univie.ac.at}
\affiliation{Department of Physics, University of Vienna, Boltzmanngasse 5, A-1090 Vienna, Austria}

\date{\today}

\begin{abstract}
 
The 3+1-dimensional Einstein-Hilbert action is obtained from the 1-loop effective action 
on noncommutative branes in the  IIB or IKKT matrix model.
The presence of compact fuzzy extra dimensions $\cK$ as well as maximal supersymmetry of the  model is essential.
The E-H action can be interpreted as interaction of $\cK$ with the 
space-time brane via IIB supergravity, and 
the effective Newton constant is determined by the Kaluza-Klein scale of $\cK$.
The classical matrix model defines a pre-gravity action with 2 derivatives less 
than the induced E-H action, governing the cosmological regime. 
The perturbative physics is confined to the space-time brane, 
which for covariant quantum space-times includes
all dof of gravity, as well as a tower of higher-spin modes.
The vacuum energy of the background is given in terms of the symplectic volume form, and hence does not 
act as cosmological constant.

\end{abstract}

\pacs{98.80.Cq}
\maketitle

\section{Introduction}

Classical gravity is well described by general relativity (GR), which arises from 
the Einstein-Hilbert action. 
However, this formulation is not well suited for quantization. Among the many approaches to 
reconcile quantum mechanics with gravity, string theory is perhaps distinguished by a sort of structural 
uniqueness. However, it leads to a vast ``landscape'' of possible compactifications to 
$3+1$ dimensions. 
In this letter, we propose a possible resolution of this problem based on a certain matrix model,
which was proposed as a constructive definition of string theory.

Our starting point is the maximally supersymmetric IIB or IKKT matrix model
\cite{Ishibashi:1996xs}
\begin{equation}
S = \frac 1{g^2}{\rm Tr}\big( [Y^{\dot\a},Y^{\dot\b}][Y_{{\dot\a}},Y_{{\dot\b}}] 
\,\, + \overline\Psi \Gamma_{\dot\a}[Y^{\dot\a},\Psi] \big) \ .
\label{MM-action}
\end{equation}
Dotted indices transform under a global $SO(9,1)$, and  are 
raised and lowered  with  $\eta^{\dot\a\dot\b}$.
Here $Y^{\dot\a} , \ \dot\a = 0,...,9$ are hermitian matrices 
or operators acting on some 
(finite-dimensional or separable) Hilbert space $\cH$, while $\Psi$ is a matrix-valued Majorana-Weyl spinor of 
$SO(9,1)$.
We will mostly ignore the fermions $\Psi$, although their presence is crucial for 
quantization. 

The action \eq{MM-action} is known to have a variety of critical points 
or backgrounds $Y^{\dot\a}$, which can  be interpreted as branes
in target space $\R^{9,1}$ carrying some $B$ field. 
The basic examples are flat branes described by Moyal-Weyl quantum planes 
$\R^{2n}_\theta \subset \R^{9,1}$, where $[Y^{\dot\a},Y^{\dot\b}] = i\theta^{{\dot\a}{\dot\b}}\one$.
Fluctuations $Y^{\dot\a} \to  Y^{\dot\a} + \cA^{\dot\a}$ of such backgrounds
are  governed by a non-commutative gauge theory \cite{Aoki:1999vr}, 
where $\cA^{\dot\a}$ becomes a gauge field
on the  brane. We focus on backgrounds leading to a weakly interacting
gauge theory. However, we  consider more general backgrounds $Y^{\dot\a}$
interpreted as quantized embedding functions  \cite{Stein2}
 \begin{align}
 Y^{\dot\a} \sim y^{\dot\a}:\quad \cM \hookrightarrow \R^{9,1}
\end{align}
of a Poisson manifold $\cM$ in target space $\R^{9,1}$, 
reducing to classical functions $y^{\dot\a}$ in the 
semi-classical (i.e. Poisson) limit. 
The $Y^{\dot\a}$ will generically not commute, and their commutator 
$[Y^{\dot\a},Y^{\dot\b}]\sim i\{y^{\dot\a},y^{\dot\b}\}$ is interpreted as quantized Poisson bracket on $\cM$. This will be considered as a (non-commutative) brane.

The quantization of this model is defined by the ``matrix path integral''
\begin{align}
 \langle Y ... Y\rangle = \frac{1}{\cZ} \int dY d\Psi Y...Y e^{i S}\qquad
  \cZ = \int dY d\Psi e^{i S} \ .
  \nn
\end{align}
This oscillatory integral becomes absolutely convergent for finite-dimensional $\cH$ upon implementing the regularization 
\begin{align}
 S\to S + i\varepsilon \sum_{\dot\a} Y_{{\dot\a}} Y_{{\dot\a}} \ ,
 \label{Feynman}
\end{align}
which amounts to a Feynman $i\varepsilon$ term in the noncommutative gauge theory.
For block-matrix configurations describing several  branes, the 
path integral leads to interactions of the branes consistent with 
IIB supergravity \cite{Ishibashi:1996xs,Chepelev:1997av,Steinacker:2016nsc}
in target space $\R^{9,1}$, thus providing a direct link with string theory.

We will focus on the perturbative physics around 
matrix configurations describing a single  noncommutative brane $\cM$. 
Then the matrices $\End(\cH) \sim \cC(\cM)$ can be interpreted as quantized functions on $\cM$.
Fluctuations around the background can thus be interpreted as fields on 
the brane, governed by a non-commutative gauge theory. 
In particular, the model leads to  $3+1$ dimensional physics on $\cM^{3,1}$ branes.
There are no fields propagating in transversal directions, in contrast to 
standard string theory.

The propagation of all fluctuations on such a $\cM^{3,1}$ brane is governed by a universal dynamical metric
specified below. 
We will demonstrate that the dynamics of this geometry is governed by an 
Einstein-Hilbert term which arises in the  one loop effective action of the IKKT model, 
{\em assuming} the presence of fuzzy 
extra dimensions $\cK$.  More specifically, we assume
a background brane with product structure\footnote{
Such backgrounds are
known to be solutions of the classical matrix model amended with
quadratic and cubic terms \cite{Aschieri:2006uw,Chatzistavrakidis:2011gs}. We assume that these terms can be replaced by
loop corrections, as suggested by \eq{V-K} and \cite{Steinacker:2015dra}.}
\begin{align}
 \cM^{3,1} \times \cK \ \subset \R^{9,1} \ ,
 \label{product-ansatz}
\end{align}
where $\cK$ is a quantized compact symplectic space 
(such as a fuzzy sphere) supporting finitely many dof, 
cf. \cite{Aschieri:2006uw,Chatzistavrakidis:2011gs}.

\section{Kinematical setup}

We consider matrix configurations $Y^{\dot\a} = (Y^{\dot a} ,Z^i)$
which describe a  noncommutative brane $\cM \times \cK \subset \R^{9,1}$ 
embedded in target space. Here
\begin{align}
 Y^{\dot a} \sim y^{\dot a} :\quad \cM \hookrightarrow \R^{3,1}, \qquad {\dot a} =0,...,3 \ 
\end{align}
describes a quantized 3+1-dimensional space-time $\cM$ embedded along the first 4 coordinate
directions, and 
\begin{align}
 Z^i \sim z^i : \quad\cK \hookrightarrow \R^{6}, \qquad i=4,...,9 \ 
\end{align}
describes a compact quantum space 
$\cK$ embedded along the transversal directions.
Dotted Latin indices $\dot a,\dot b =0,...,3$ 
 indicate frame-like indices transforming under the global 
$SO(1,3)$ symmetry of the  model.
The Hilbert space is assumed to factorize
as $\cH = \cH_\cM \otimes \cH_\cK$,
where $\cH_\cK = \C^n$ corresponds to the compact  space $\cK$.
The matrix d'Alembertian $\Box = [Y^{\dot \a},[Y_{\dot \a},.]]$ then decomposes as 
\begin{align}
 \Box = [Y^{\dot a},[Y_{\dot a},.]] + [Z^i,[Z_i,.]]
 = \Box_\cM + \Box_\cK \ .
\end{align}
The detailed structure of $\cK$ will be irrelevant\footnote{in the simplest case $\cK$ could be a fuzzy sphere $S^2_N$, 
and more interesting spaces leading to interesting low-energy gauge theories 
have been considered, cf. \cite{Chatzistavrakidis:2011gs,Sperling:2018hys}.}.
We only require that the internal 
matrix Laplacian $\Box_\cK$ has a positive spectrum, 
\begin{align}
 \Box_\cK \l_{\L} = m^2_\L\, \l_{\L} \ , \qquad m_{\L}^2 = m_\cK^2 \mu^2_{\L} \ 
 \label{KK-masses}
\end{align}
with eigenmodes
$ \l_{\L} \in \End(\cH_\cK)$ enumerated by some discrete label $\L$.
We then expand the most general  mode $\phi\in \End(\cH)$
into the finite basis of Kaluza-Klein (KK) modes 
\begin{align}
 \phi_{\L} = \phi_{\L}(x) \l_{\L}
 \label{phi-product-extra}
\end{align}
with $\phi_{\L}(x) \in \End(\cH_\cM)$,
which satisfy
\begin{align}
 \Box \phi_{\L} = (\Box_\cM + m_{\L}^2) \phi_{\L} \ .
\end{align}
Therefore the $\phi_{\L}$ mode acquires a mass $m_{\L}^2$ on $\cM$,
as in standard KK compactification.

Now consider the quantum space-time $\cM$ defined by $Y^{\dot a}$. 
To understand the kinematics, we consider the semi-classical regime
where commutators $[.,.] \sim i\{.,.\}$ can be replaced by Poisson brackets, and define
\begin{align}
 \Theta^{\dot a\dot b} = -i[Y^{\dot a},Y^{\dot b}]
 \sim  \{Y^{\dot a},Y^{\dot b}\} \ .
 \label{hat-Theta-def-2}
\end{align}
Then the matrices $Y^{\dot a}$ define a frame as follows 
\begin{align}
 E^{\dot a \mu} &:= \{Y^{\dot a},x^\mu\}  \label{frame-general}
\end{align} 
where $x^\mu$ are local coordinates on $\cM$.
All fluctuations in the matrix model are governed by a kinetic term of the structure
\begin{align}
  \Tr([Y^{\dot a},\Phi][Y_{\dot a},\Phi]) \sim \int d^4x\sqrt{|G|}\, G^{\mu\nu}\del_\mu\phi\del_\nu\phi
\end{align}
in the semi-classical limit $\Phi \sim \phi$. This defines the 
{\bf effective metric} on $\cM$:
\begin{align}
G^{\mu\nu} &= \r^{-2}\, \g^{\mu\nu} , \qquad 
 \g^{\mu\nu} = \eta_{{{\dot a}}{\dot b}} \tensor{E}{^{\dot a}^\mu} \tensor{E}{^{\dot b}^\nu} \ ,
  \label{eff-metric-def}
\end{align} 
for a uniquely determined dilaton $\rho$ \cite{Stein2}.
To capture the geometry in the matrix model, it is useful to consider the following
``torsion'' tensor
\begin{align}
 \tensor{T}{^{\dot a}^{\dot b}^\mu} &:= -\{\Theta^{\dot a\dot b},x^\mu\} 
  = -\tensor{E}{^{\dot a}^\nu}\del_\nu\tensor{E}{^{\dot b}^\mu} 
  + \tensor{E}{^{\dot b}^\nu}\del_\nu\tensor{E}{^{\dot a}^\mu}
 \label{torsion-general}
\end{align}
using  the Jacobi identity.
This can be understood as torsion tensor of the Weizenb\"ock 
connection $\nabla^{(W)} E^{\dot a} = 0$ \cite{Steinacker:2020xph}, cf. \cite{Langmann:2001yr}.
We replace the frame indices with coordinate indices by contracting with 
 coframes 
$\tensor{E}{_{\dot a}_\s}\tensor{E}{_{\dot b}_\k}$
defined via $\tensor{E}{^{\dot a}^\mu}\tensor{E}{_{\dot a}_\nu} = \d^\mu_{\nu}$, 
and obtain
\begin{align} 
 \tensor{T}{_\s_\k^\mu} 
  = \tensor{E}{^{\dot a}^\mu}
   (\del_\s\tensor{E}{_{\dot a}_\k} - \del_\k\tensor{E}{_{\dot a}_\s}) \ .
   \label{T-dE}
 \end{align}
 This becomes more transparent in terms of the 2-form 
 \begin{align}
 T^{\dot a} =  \frac 12\tensor{T}{_\mu_\nu^{\dot a}} dx^\mu dx^\nu = d \theta^{\dot a}
 \label{T-dtheta}
 \end{align}
 where 
\begin{align}
 \theta^{\dot a} := \tensor{E}{^{\dot a}_\mu} dx^\mu  
\end{align}
 is the coframe one-form. 
Therefore the torsion 2-form is nothing but the exterior derivative 
of the coframe. 
In the matrix model framework, the first form in \eq{torsion-general} is most relevant;
for a more detailed discussion see \cite{Steinacker:2020xph,Fredenhagen:2021bnw}.

\paragraph{Covariant quantum spaces.}

For 3+1-dimensional noncommutative
branes $\cM$, the tangential  fluctuations cannot provide the 
most general metric dof. Transversal fluctuations are presumably suppressed in the presence of $\cK$.
Furthermore, the presence of an antisymmetric tensor field $\theta^{\dot a\dot b}$ on space-time is problematic.  
These issues are resolved for covariant quantum spaces, 
 which are twisted bundles over space-time with a 2-dimensional fuzzy sphere $S^2_n$ as fiber  \cite{Sperling:2019xar,Steinacker:2017bhb,Heckman:2014xha}, 
 \begin{align}
  \cM  \stackrel{loc}{\cong} \cM^{3,1} \times S^2_n \ 
  \label{cov-bundle}
 \end{align}
embedded in the matrix model via 3+1 generators $Y^{\dot\a}$. For example, a FLRW space-time is
realized by 
\begin{align}
 Y^{\dot\a} = \frac{1}{R}\cM^{\dot a 4}
 \label{cov-spacetime}
\end{align}
acting on a doubleton irrep $\cH_n$ of $\mso(4,2)$.
 One can then expand the fluctuations into harmonics $\phi_{sm} = \phi_{sm}(y) Y_{sm}$ on $S^2_n$, 
 so that \cite{Sperling:2019xar}
\begin{align}
 \Box_\cM = \Box_{\cM^{3,1}}  +  m^2_s, \qquad m_s^2 = \frac{s(s-1)}{R^2} \ .
\end{align}
Here $R$ is a  large (cosmological) scale parameter. 
Due to the twisted bundle structure,
the $Y_{sm}$ turn out to be spin $s$ modes on space-time. This
leads to a {\em finite}  tower of higher-spin excitations for $s\leq n$,
which provides all degrees of freedom of gravity in 3+1 dimensions \cite{Sperling:2019xar}.
The above discussion for frame and torsion generalizes easily, 
in terms of higher-spin valued frame and torsion.
We will ignore their higher-spin contributions
in the following, and focus on the
geometric sector.

\section{One-loop effective action}

We wish to compute the 1-loop effective action on a background of the above type, 
in the weak coupling regime.
We first ignore the internal fiber $S^2_n$ of covariant quantum spaces,
which will be included later.
The one-loop  effective action of the  IKKT model for some given matrix background 
$Y^{\dot \a}$ is defined by
\begin{align}
 Z_{\rm 1\, loop}[Y] &= \int\limits_{\rm 1\, loop} dY d\Psi 
 e^{iS[Y,\Psi]} 
 = e^{i (S[Y] + \Gamma_{\!\rm{1 loop}}[Y]) } 
 \nn
 \end{align}
in terms of an oscillatory Gaussian integral around  $Y$.
Taking into account the fermions and the ghost contributions,
this leads to \cite{Ishibashi:1996xs,Chepelev:1997av,Blaschke:2011qu} 
\begin{align}
\Gamma_{\!\textrm{1loop}}[Y]\! &= \frac i2 \Tr \Big(\log(\Box + M^{(V)}_{\dot\a\dot\b}[ \Theta^{\dot\a\dot\b},.]) \nn\\
&-\frac 12 \log(\Box + M^{(\psi)}_{\dot\a\dot\b}[ \Theta^{\dot\a\dot\b},.] )
- 2 \log\Box\Big)\ .
\label{Gamma-IKKT}
\end{align}
 Here  the trace is over hermitian matrices in $\End(\cH)$, 
  and
\begin{align}
\begin{array}{rl}M_{\dot\a\dot\b}^{(\psi)} &= \frac 1{4i} [\Gamma_{\dot\a},\Gamma_{\dot\b}] \,  \\
(M_{\dot\a\dot\b}^{(V)})^{\dot\g}_{\dot\d} &= i(\d^{\dot\g}_{\dot\b} \eta_{\dot\a\dot\d} - \d^{\dot\g}_{\dot\a} \eta_{\dot\b\dot\d}) \, , \\
                    \end{array}  
\end{align}
are  $SO(9,1)$ generators acting on the vector or spinor representation, respectively.
The  $2 \log \Box$ term arises from the ghost contribution.
Using the expansion 
\begin{align}
 \log(\Box  + \cO)
  = \log\Box + \sum_{n>0} \frac{1}n \big(\Box^{-1}\cO \big)^n \ 
\end{align}
and observing that 
the first 3 terms in this expansion cancel  due to maximal supersymmetry,
the leading non-trivial term is the 4th order term
given by \cite{Blaschke:2011qu}
\begin{align}
 \Gamma^{\textrm{1loop}}_{4}\! &=
 \frac i8 \Tr \Big(\!  \big(\Box^{-1}M^{(V)}_{\dot\a\dot\b} [\Theta^{\dot\a\dot\b},.]\big)^4 \!
  -\frac 12 \big(\Box^{-1}M^{(\psi)}_{\dot\a\dot\b} [\Theta^{\dot\a\dot\b},.]\big)^4 \Big)
\label{Gamma-IKKT-4}
\end{align}
dropping $O(\Box^{-5})$ contributions.
In  these expressions, $\Box \to \Box -i\varepsilon$ arising from 
\eq{Feynman} is understood.

\paragraph{Evaluation of the trace.}

We will evaluate the trace over $\Tr_\cM \equiv \Tr_{\End(\cH_\cM)}$ 
 using the basic formula \cite{Steinacker:2016nsc}
\begin{align}
 \Tr_{\cM} \cO &= \frac{1}{(2\pi)^{4}}\,\int\limits_{\cM\times\cM} \Omega_x \Omega_y
   \left(^x_{y} \right| \cO \left|^x_{y} \right)
  \label{trace-coherent-End}
\end{align}
where $\Omega = d^4 x \r_\cM$ is the symplectic volume form.
Here 
\begin{align}
\left|^x_{y} \right) 
 &:= |{y}\rangle\langle {x}| \ \ \in \End(\cH_\cM)
\end{align}
 are string modes in terms of coherent states $|{x}\rangle$
on the symplectic space $\cM$. 
This formula is exact for homogeneous quantum spaces, and follows
from group invariance \cite{Steinacker:2016nsc}. 
As a symplectic manifold, we can assume that $\cM$ is equivalent to a 4-dimensional quantized homogeneous space
such as $\R^{4}_\theta$; the extra $S^2_n$ factor on covariant quantum spacetime 
will be taken into account later.

String modes are  useful to evaluate the trace over $\End(\cH_\cM)$, because they
enjoy approximate localization properties in {\em both} position and momentum
\cite{Steinacker:2016nsc,Iso:2000ew}.
In particular, 
\begin{align}
 \Box(| {y}\rangle\langle {x}|) &\sim (|x- y|^2 + L_{NC}^2) \ | {y}\rangle\langle {x}|  \nn\\
 [ \Theta^{\dot a\dot b},.] (| {y}\rangle\langle {x}|) 
 &\sim  \d\Theta^{\dot a\dot b}( {y},{x}) | {y}\rangle\langle {x}| \nn\\
 \d\Theta^{\dot a\dot b}( {y},{x}) &=  \Theta^{\dot a\dot b}( {y})-  \Theta^{\dot a\dot b}({x}) \ .
 \label{string-modes-EV}
\end{align}
where $|x-y|$ is the distance in target space $\R^{9,1}$, and
\begin{align}
 L_{NC}^2 = \det(\theta^{\dot a\dot b})^{1/4}
\end{align}
is the scale of noncommutativity.
Thus $\Box$  measures the length of the string modes, which are responsible for UV/IR mixing
 in non-SUSY models \cite{Minwalla:1999px}.

We can therefore  evaluate the 1-loop integral approximately as follows
\begin{align}
\Gamma^{\textrm{1loop}}_4\! 
  &=  \frac i4 \frac{1}{(2\pi)^{4}}\,\Tr_{\cK}\!\!\!
  \int\limits_{\cM\times\cM} \Omega_x \Omega_y
   \frac{3 V_4[ \d\Theta({x},{y})] }{(|x-y|^2+L_{NC}^2)^4} 
\label{1-loop-coh}
\end{align}
where \cite{Steinacker:2016nsc}
\begin{align}
 V_4[ \d\Theta]  &= \d\Theta^{\dot \a\dot \b} \d\Theta^{\dot \a\dot \b}\d\Theta^{\dot \g\dot \d} \d\Theta^{\dot \g\dot \d}
  -4\d\Theta^{\dot \a\dot \b} \d\Theta^{\dot \b\dot \g}\d\Theta^{\dot \g\dot \d}\d\Theta^{\dot \d\dot \a} \nn\\
   &= -4tr (\d\Theta^4) + (tr \d\Theta^2)^2 \ .
 \label{S-4-short}
\end{align}
Here $\d\Theta^{ij} \to [\Theta^{ij},.]$ is understood
for the contributions of $\cK$.
On a product space $\cM\times\cK$,
this leads to the following contributions 
 \begin{align}
 V_4 &= V_4^\cM +  V_4^\cK  +  V_4^{\cM\cK} \ .
 \label{V4-decomp}
\end{align}
The most interesting term is 
\begin{align}
  V_4^{\cM\cK} &= 2\d\Theta^{\dot a\dot b} \d\Theta^{\dot a\dot b}
     [\Theta^{ij},[\Theta^{ij},.]] \ 
 \label{S4-explicit-M-K}
\end{align}
which is responsible for the interaction between $\cM$ and $\cK$, and gives rise to  
a gravity action on $\cM$.
There are no other contributions, since the mixed components of the Poisson tensor 
$[Y^{\dot a},Y^i] =0$ vanish.
The contribution of $\cK$ in \eq{Gamma-IKKT-4} can be evaluated for the product states 
\eq{phi-product-extra}, noting that
\begin{align}
 [\Theta^{ij},\phi_{\L}(y) \l_{\L}]
  &= \phi_{\L}(y)[\Theta^{ij},\l_{\L}] \ .
\end{align}
We assume  for simplicity that 
  $\l_{\L}$ is a common eigenvector of both $\Box_\cK$
and $[\Theta^{ij} ,[\Theta^{ij},.]] $ in $\End(\cH_\cK)$.
Then
\begin{align}
 [\Theta^{ij} ,[\Theta^{ij},\l_{\L}]] = m_\cK^4 C^2_{\L} \l_{\L} \ , 
 \label{K-mixed-contrib}
\end{align}
where $m_\cK^2$ is the KK mass scale, and $C^2_{\L}>0$ are  
numerical constants depending on the structure of $\cK$.

To compute $V_4^{\cM\cK}$, we need to evaluate 
$\big(\d\Theta^{\dot a\dot b} \d\Theta^{\dot a\dot b}\big)$ on the string states, recalling that $\d\Theta^{\dot a\dot b}$
stands for $[\Theta^{\dot a\dot b},.]$. 
Since the integral \eq{1-loop-coh} is convergent due to the $\frac{1}{|x-y|^8}$ behavior,
only ``short'' string states with $|x-y| \leq L_{NC}$ contribute,
which capture the low-energy sector of $\End(\cH_\cM)$.
Then $[\Theta^{\dot a\dot b},.] \sim i\{\Theta^{\dot a\dot b},.\}$ can be evaluated in the 
semi-classical regime, but the approximate relations \eq{string-modes-EV} need to be refined.
This is achieved by recognizing that short string states correspond
precisely to localized Gaussian wave packets on  $\cM \approx\R^4_\theta$:
 \begin{align}
 e^{\frac i2 k^\mu \theta^{-1}_{\mu\nu} y^\nu}\big|^{y+\frac k2}_{y-\frac k2} \big)  
 =: \Psi_{\tilde k;y}  \cong  \psi_{\tilde k;y}(x) = \frac{2}{\pi L_{NC}^2}\, e^{i \tilde k x} e^{-|x-y|_g^2} \ .
\nn  
 \end{align} 
Here $\tilde k_\mu = k^\nu\theta^{-1}_{\nu\mu}$, and
 $|.|_g = L_{NC}^{-1} |.|$ is the metric which renders $(\cM,\omega,g)$
 an almost-K\"ahler manifold.
This provides an isometric identification of the low-energy modes in $\End(\cH_\cM)$ and $L^2(\cM)$, which
can be checked explicitly  using the inner product of coherent states 
\begin{align}
 \langle x|y\rangle &= e^{\frac i2 y^\mu \theta^{-1}_{\mu\nu} x^\nu} e^{-\frac 14 |x-y|^2_g} \ .
  \label{inner-coh-general}
\end{align}
Hence $\psi_{k;y}$ are Gaussian wave packets of characteristic size $L_{NC}$ centered at $y$. 
More abstractly,
we can identify the space of short string states  with the cotangent bundle
\begin{align}
 (\cM \times \cM)_{\rm loc} &\cong T^*\cM  \nn\\
  (x,y) &\mapsto (x,\tilde k), 
  \qquad \tilde k_\nu = (y-x)^\mu \theta^{-1}_{\mu\nu} \ .
 \nn
\end{align}
The symplectic  measure on $\cM\times\cM$ then reduces to the canonical 
measure on $T^*\cM$.
This allows to rewrite the trace formula \eq{trace-coherent-End} as follows
\begin{align}
  \Tr_\cM \cO 
 &\approx \frac{1}{(2\pi)^{4}}\!\int_\cM dx \int d k
  \langle \Psi_{k,x}, \cO  \Psi_{k,x}\rangle \nn\\
 &=  \frac{1}{(2\pi)^{4}}\!\int\limits_{\cM}  \sqrt{G} dx \int \frac{d k}{\sqrt{G}}
  \langle \psi^{(L)}_{k,x}, \cO  \psi^{(L)}_{k,x}\rangle
  \label{trace-NC-conv-int}
\end{align} 
which is independent of $L$.
The second formula  applies also in the commutative case, and
 \begin{align}
  \psi^{(L)}_{k;y} &:= 
   \int d^4z\, e^{-|y-z|_g^2 \frac{L_{NC}^2}{L^2}}\,\psi_{k;z}
  \label{smeared-string}
 \end{align}
are Gaussian averages of the short string modes with size $L \gg  L_{NC}$.
These are approximately plane waves $\psi_{k} \propto e^{ikx}$,
which allows to evaluate $[\Theta^{\dot a\dot b},.] \sim i\{\Theta^{\dot a\dot b},.\}$
as 
\begin{align}
 \{\Theta^{\dot a\dot b},\psi_{k}\} &= \{\Theta^{\dot a\dot b},x^\mu\}\del_\mu\psi_{k}
 \approx - i T^{\dot a\dot b\mu} k_\mu \psi_{k} \nn\\
 \{\theta^{\dot a\dot b}, \{\theta^{\dot a\dot b},\psi_{k;y}\}\} 
 &\approx -\{\theta^{\dot a\dot b},x^\mu\}\{\theta^{\dot a\dot b}, x^{\nu} \} k_\mu  k_{\nu}\psi_{k;y} \nn\\
   &= -\tensor{T}{^{\dot a}^{\dot b}^\mu} \tensor{T}{^{\dot a}^{\dot b}^{\nu}} k_\mu k_{\nu}\psi_{k;y}\ .
 \nn 
\end{align}
This is justified if the torsion $T^{\dot a\dot b\mu}$ \eq{torsion-general} is approximately constant
on length scales $L$, which can be assumed in the context of gravity.
Putting this together,
 $V_4^{\cM\cK}$ acting on the $\psi_{k;y}\l_{\L}$ is 
 approximately diagonal, and reduces to 
\begin{align}
 V_4^{\cM\cK} [\psi_{k;y}\l_{\L}]
 = 2 m_\cK^4 C^2_{\L} \tensor{T}{^{\dot a}^{\dot b}^\mu} \tensor{T}{^{\dot a}^{\dot b}^{\nu}} k_\mu k_{\nu}
 \ \psi_{k;y}  \l_{\L} \ . \nn 
\end{align}
Hence the full trace reduces to the following local integral 
\begin{align}
 &\Gamma_{\rm 1 loop}^{\cK-\cM}  =
  \frac{3i}{4}\Tr\Big(\frac{V_4^{\cM\cK}}{(\Box-i\varepsilon)^4}\Big) \nn\\
  &= \frac{3 i m_\cK^4}{2(2\pi)^4}\int\limits_{\cM}d^4 x\sqrt{G} \sum_{\L} C^2_{\L} \int \frac{d^4 k}{\sqrt{G}}\,
  \frac{\tensor{T}{^\a^\b^\mu}  \tensor{T}{^\a^\b^{\nu}} k_\mu k_{\nu}}{(k\cdot k + m^2_{\L}-i\varepsilon)^4} 
  \nn
 \end{align}  
 where $k\cdot k = k_\m k_\nu \g^{\mu\nu}$,
 re-inserting the $i\varepsilon$. 
 The integral over $k$ can be evaluated using contour integration as 
  \begin{align}
   \int \frac{d^4 k}{\sqrt{G}}\, \frac{k_\mu  k_{\nu}}{(k\cdot k + m^2 - i \varepsilon)^4} \
  &= \frac{i\pi^2}{12m^2} \r^{-6} G_{\mu\nu} \ .
\end{align}
 Since the frame $E^{\dot\a}_\mu$ and the torsion correspond to the 
metric $\g^{\mu\nu} = \r^2 G^{\mu\nu}$ \eq{eff-metric-def}, it follows that
\begin{align}
  \r^{-4} \tensor{T}{^{\dot\a}^{\dot\b}^{\mu}}\tensor{T}{^{\dot\a}^{\dot\b}^{\nu}} G_{\mu\nu}
    = \tensor{T}{^\r_\s_{\mu}}\tensor{T}{_{\r}^{\s}_{\nu}} G^{\mu\nu} \ .
\end{align}
Therefore 
\begin{align}
 \Gamma_{\rm 1 loop}^{\cK-\cM}  
 = -\frac{c_{\cK}^2}{(2\pi)^4} \int\limits_{\cM}d^4 x\sqrt{G}\, \r^{-2} m_\cK^2
  \tensor{T}{^\r_\s_{\mu}}\tensor{T}{_{\r}^{\s}_{\nu}} G^{\mu\nu} 
  \label{T-T-action}
\end{align}
where 
 \begin{align}
 c_{\cK}^2 = \frac{\pi^2}{8}\sum_{\L} \frac{C^2_{\L}}{\mu_{\L}^2}\  > 0 \ 
 \label{C2-K-cutoff}
\end{align}
is finite, determined by the dimensionless KK masses \eq{KK-masses}.
This is recognized as gravitational action using the following identity 
obtained from (E.2)  in \cite{Fredenhagen:2021bnw}
\begin{align}
  \cR &=  - \frac 12\tensor{T}{^\mu_\s_\r} \tensor{T}{_\mu_{\s'}^\r} G^{\s\s'}
  - \frac 12 T_{\nu} T_{\mu}  G^{\mu\nu}  
        + 2 \r^{-2} G^{\mu\nu}\del_\mu\r\del_\nu\r \nn\\
   &\quad     - 2\nabla_{(G)}^\mu (\r^{-1} \del_\mu\r) \ .
 \nn  
\end{align}
Here $\cR$ is the Ricci scalar of the effective metric $G_{\mu\nu}$, and
$T_{\mu} dx^\mu = -\star(\frac 12 G_{\nu \s}\tensor{T}{^{\s}_\r_\mu}dx^\nu dx^\r dx^\mu)$
 is the Hodge-dual
of the totally antisymmetric torsion, which reduces using the 
 eom of the matrix model to a gravitational axion $\tilde\r$  \cite{Fredenhagen:2021bnw}
\begin{align}
 T_\mu = \r^{-2}\del_\mu\tilde\r \ .
\end{align}
Therefore we obtain the Einstein-Hilbert action with an extra contribution from $T_\mu$ and $\rho$:
\begin{align}
 \Gamma_{\rm 1 loop}^{\cK-\cM}
 &=  \int\limits_\cM \! d^{4}x\frac{\sqrt{|G|}}{16 \pi G_N}\,
   \Big(\cR
  + \frac 12 T_{\nu} T_{\mu}  G^{\mu\nu}  
      \!  - 2 \r^{-2} \del_\mu\r\del^\mu\r \nn\\
  &   \qquad   + 2 \r^{-1} \del_\mu\r\, G_N^{-1}\del^\mu G_N \Big) \ .
   \label{E-H}
\end{align}
%
%
%
with  Newton constant set by
 the compactification scale 
 \begin{align}
 \frac{1}{G_N} = \frac{2c^2_{\cK}}{\pi^3} \r^{-2} m_\cK^2\, .
 \end{align}
 Hence the Planck scale is related to the Kaluza-Klein scale for the 
 fuzzy extra dimensions $\cK$, which also serves as an effective UV cutoff.
 Since the 1-loop effective action is related to IIB supergravity,
 the gravity {\em action} in 3+1 dimensions can be interpreted as  quasi-local {\em interaction}
of $\cK$ and $\cM$ via 9+1-dimensional IIB supergravity.

It turns out that
$\Gamma_{\rm 1 loop}^{\cK-\cM} = c^2 m_\cK^2 > 0$ \eq{T-T-action} is positive for the covariant 
FLRW space-time in \cite{Sperling:2019xar}. 
Combined with the bare action, the effective potential has the structure 
\begin{align}
 V(m_\cK^2) = - c^2 m_\cK^2 + \frac{d^2}{g^2} m_\cK^4
 \label{V-K}
\end{align}
 at weak coupling,
which has a minimum for $m_\cK^2>0$. 
Since $m_\cK$ is set by the radius of $\cK$,
this indicates that  $\cK$ can be stabilized by quantum effects,  
providing some justification for \eq{product-ansatz}.

\paragraph{Vacuum energy.}

The vacuum energy  arising at 1 loop from 
$\cK$  is obtained using an analogous trace computation, 
leading to a result of  structure 
\begin{align}
\Gamma_{\rm 1 loop}^\cK
 = \frac{3i}4\Tr\Big(\frac{V_4^\cK}{\Box^4}\Big)
  &\sim -\frac{\pi^2 }{8(2\pi)^4}
  \int\limits_\cM \!\Omega\, \r^{-2} m_\cK^4\sum_{\L s}
  \frac{V_{4,\L}}{\mu^4_{\L}}  
  \label{vacuum-energy-K}
\end{align}
assuming $\frac 1{R^2} \ll m^2_{\L}\sim m_\cK^2$. 
Here $V_{4,\L}$ is determined by the structure of $\cK$, and
could  have either sign.
Since the symplectic volume form  $\Omega$ 
is independent of the metric,
the 1-loop vacuum energy does not act as a cosmological constant.
The present setup can therefore be viewed as a realization of induced
gravity in the spirit of Sakharov \cite{Sakharov:1967pk,Visser:2002ew}, 
avoiding the associated cosmological constant problem.
Finally,  $V_4^\cM$ \eq{V4-decomp} leads to a 4-derivative contribution 
$\Gamma_{\rm 1 loop}^\cM$,
which is expected to be negligible for long wavelengths.

\paragraph{Covariant quantum space-time.}

The above computation  extends
straightforwardly to covariant quantum space-times \eq{cov-bundle}, leading again to
the gravity action \eq{E-H}, with $c_{\cK}$ modified as
\begin{align}
  c_{\cK}^2 = \frac{\pi^2}{8}\sum_{\L s} \frac{(2s+1)C^2_{\L}}{\mu_{\L}^2 + \frac{s(s-1)}{R^2 m^2_\cK}}\  
  \approx  \frac{\pi^2}{8}\sum_{\L s} \frac{(2s+1)C^2_{\L}}{\mu_{\L}^2 }
 \label{C2-K-cutoff-cov}
\end{align}
assuming $R^2 m^2_\cK \gg 1$.
The vacuum energy \eq{vacuum-energy-K} is also  slightly modified with the 
same qualitative features, and does again not gravitate. 
More details will be given in a forthcoming publication.

\section{Discussion}

We have shown that the  3+1-dimensional Einstein-Hilbert action arises at one loop in the
IKKT matrix model on suitable 3+1-dimensional banes, in the 
presence of fuzzy extra dimensions $\cK$ but without target space compactification. 
The vacuum energy is large but does not gravitate,
due to the symplectic structure of the brane.
Combining the 1-loop contribution \eq{E-H} with the bare action, the effective action for gravity
up to second derivatives in the frame has the form 
\begin{align}
 S_{\rm grav} = -\int \Omega \frac 1{g^2}\Theta_{\dot a\dot b}\Theta^{\dot a\dot b}
  + \Gamma_{\rm 1 loop}^{\cK-\cM} \ .
\end{align}
The bare action has 2 derivatives less than the E-H action and
is hence interpreted as ``pre-gravity'' \cite{Steinacker:2016vgf,Fredenhagen:2021bnw}, which should dominate at long (cosmic)
scales. At shorter scales the E-H term will dominate, leading to a cross-over with GR.
The mechanism works only in the maximally supersymmetric model, where the gauge theory
on the brane is UV finite. 
The induced E-H action  on $\cM$ can be understood as quasi-local
IIB supergravity interaction between $\cK$ and $\cM$ with negative binding energy,
suggesting that such configurations can be stable.
It should be possible  to verify (meta)stability  of
such configurations by numerical simulations in the matrix model,
but this is very challenging, cf.  \cite{Nishimura:2019qal,Anagnostopoulos:2020xai} 
for related numerical work.

Finally, the presence of  $\cK$ leads to a non-trivial gauge theory on $\cM$,
and fermions coupling to chiral gauge theories can arise 
from self-intersections of $\cK$ \cite{Chatzistavrakidis:2011gs,Aoki:2014cya,Sperling:2018hys}. 
This may lead to physically interesting gauge theories coupled to 
the above ``emergent'' gravity in a consistent quantum framework.

\section*{Acknowledgments}

A related collaboration with J. Tekel 
and useful discussions with V.P. Nair are 
gratefully  acknowledged.
This work was supported by the Austrian Science Fund (FWF) grant P32086.

\end{document}